\documentclass[useAMS,usegraphicx]{mn2e}

\def\H2{H$_2$}
\def\nH{n_{\rm H}}
\def\fH2{f_{\rm H_2}}
\def\j21{J_{21}}

\def\ie{{i.e.\ }}
\def\eg{{e.g.\ }}

\begin{document}
\title[H$_2$ in damped Ly$\alpha$ systems]{Molecular Hydrogen in
Damped Ly$\alpha$ Systems: Spatial Distribution}
\author[H. Hirashita et al.]{H. Hirashita$^{1}
\thanks{e-mail: irasita@arcetri.astro.it}$\thanks{Postdoctoral
     Fellow of the Japan Society for the Promotion of
     Science (JSPS) for Research Abroad},
A. Ferrara$^{2}$, K. Wada$^{3}$, and P. Richter$^{1}$
\\
$^1$ Osservatorio Astrofisico di Arcetri, Largo Enrico Fermi, 5,
     50125 Firenze, Italy \\
$^2$ SISSA/International School for Advanced Studies, Via
     Beirut 4, 34014 Trieste, Italy \\
$^3$ National Astronomical Observatory of Japan, Mitaka, Tokyo
     181-8588, Japan \\
}
\date{Accepted 2003 March 21}
\pubyear{2003} \volume{000} \pagerange{1--5}
\twocolumn

\maketitle \label{firstpage}
\begin{abstract}
To interpret \H2 quasar absorption line observations in Damped
Ly$\alpha$ clouds (DLAs), we model the \H2 spatial distribution
within a DLA. Based on  numerical simulations of disk structures
with parameters similar to those derived for
such absorbers, we calculate the \H2 distribution as a function of
ultraviolet background (UVB) intensity and dust-to-gas ratio.
For typical values of these two quantities we find that the area
in which the \H2 fraction exceeds $10^{-6}$ (typical
observational detection limit) only covers $\la 10$\% of the disk
surface, \ie \H2 has a very inhomogeneous, clumpy distribution
even at these low abundance levels. This explains the relative
paucity of \H2 detections in DLAs. We also show the dependence of
the covering fraction of \H2 on
dust-to-gas ratio and UVB intensity and we comment on
the physics governing the \H2 chemical network at high redshift.
\end{abstract}

\section{Introduction}

In the recent years, evidence has been found for the existence of
heavy elements in damped Ly$\alpha$ clouds (DLAs), \ie
quasar absorption line systems whose neutral hydrogen column
density is larger than $\sim 1-2\times 10^{20}$ cm$^{-2}$
(\eg Pettini et al.\ 2001; Prochaska \& Wolfe 2002). The evolution
of metal abundance in DLAs can trace the chemical enrichment
history of present galaxies. Based on this, and on other
clues, DLAs have been suggested to be the progenitors of nearby
galaxies; the similar values of the baryonic mass density in
DLAs around redshift $z\sim 2$ and the stellar mass density at
$z\sim 0$ has further supported this idea (Lanzetta, Wolfe,
\& Turnshek 1995).

It is natural to consider that the DLAs contain a certain amount
of dust, because dust-to-gas ratio and metallicity are
correlated even for metal-poor galaxies (Schmidt \& Boller 1993;
Lisenfeld \& Ferrara 1998). Indeed, Fall, Pei, \& McMahon (1989)
have suggested that the reddening of background quasars indicates
typical dust-to-gas ratios of $\sim 1/20$--1/4 of the Milky Way
(see also Zuo et al.\ 1997). The depletion of heavy elements
also supports the dust content in DLAs (\eg Vladilo 2002). The
existence of dust implies the possibility that the formation of
hydrogen molecules (\H2) is enhanced because of the \H2
grain surface reaction. Hirashita \& Ferrara (2002) have recently
shown that even in metal poor ($\sim 0.01~Z_\odot$) galaxies,
dust grains
can drastically accelerate the formation rate of \H2. They also
argued that the enhancement of molecular abundance
results in an enhancement of the star formation activity in
the early evolutionary stages of galaxy evolution, because
stars only form in molecular clouds. The important role of dust
on the enhancement of the \H2 abundance is also suggested by
observations of DLAs (Ge, Bechtold, \& Kulkarni 2001)
and in the local Universe, e.g., in Galactic halo clouds
(\eg Richter et al.\ 2001) and in the Magellanic Clouds
(Richter 2000; Tumlinson et al.\ 2002).

Although the \H2 fraction,
$\fH2\equiv 2N({\rm H_2})/[2N({\rm H_2})+N(\mbox{H\,{\sc i}})]$,
where $N(X)$ indicates the column density of the species $X$, is
largely enhanced for some DLAs, stringent upper limits are laid
on a significant fraction of DLAs in the range
$\sim 10^{-7}$--$10^{-5}$
(Petitjean, Srianand, \& Ledoux 2000). This
can be interpreted as due to a low formation rate of
\H2 in dust-poor environments
relative to the Milky Way (Levshakov et al.\ 2002; Liszt 2002)
and high \H2 dissociation
rate by strong ultraviolet background (UVB) radiation
(e.g., Petitjean et al.\ 2000). However, we should keep in
mind that such upper limits do not exclude the existence of
molecular-rich clouds in these systems, because molecular
clouds may have a very low volume filling factor. If the
covering fraction of molecular-rich regions on a galactic
surface is extremely small, it is natural that \H2 is hardly
detected in DLAs. Thus, the argument on the \H2 abundance in DLAs
is strongly dependent on the geometry of \H2 distribution
within those systems.

In order to get a better understanding of the spatial distribution
of \H2, which can then be used to interpret observations, we
present here a study of such problem based on high-resolution
numerical simulations. This allows us 
to tackle the problem of the \H2 formation/destruction and
distribution in DLAs in a realistic way.
We calculate the spatial structure of \H2 distribution in a
galactic disk under various conditions by varying the UVB intensity
and dust-to-gas ratios. As the underlying 
gas density and temperature distribution is, to a first 
approximation, independent of the \H2 properties (for example, it 
does not contribute to cooling at the metallicity level typical
for DLAs),
this approach offers the opportunity to explore the parameter space
with the required accuracy. 

Throughout this Letter, we assume a flat $\Lambda$CDM
cosmology\footnote{We use the following values of the cosmological
parameters: $\Omega_{\rm M}=0.3$,
$\Omega_\Lambda =0.7$, 
$H_0\equiv 100 h$ km s$^{-1}$ Mpc$^{-1}=70$ km s$^{-1}$ Mpc$^{-1}$, 
and $\Omega_{\rm b}=0.02h^{-2}$.} 
(Mo \& White 2002).
We first describe the simulation that we used to derive the
density and temperature maps (Section \ref{sec:simulation}).
The maps are presented in Section \ref{sec:spatial}, where the
\H2 distribution maps are also shown. Based on these results,
we discuss our interpretation of current observations of \H2
in DLAs (Section \ref{sec:discussion}).

\section{Numerical Simulation of DLA disks}
\label{sec:simulation}

It is still unclear whether DLAs are large protogalactic disks
(Prochaska \& Wolfe 1998), protogalactic clumps (Haehnelt,
Steinmetz, \& Rauch 1998; Ledoux et al.\ 1998), or a mixture
of various populations (Cen et al.\ 2002). Here, we assume
that DLAs are large protogalactic disks. We use a
two-dimensional hydrodynamical simulation assuming a disk-like
geometry (for the method, see
Wada \& Norman 2001).
However, the statistical properties of
density and temperature are determined by non-linear
hydrodynamical effects (Wada \& Norman 2001), and we expect that
the results of this Letter remain valid if other scenarios for
DLA formation are adopted.

We have run the hydrodynamical calculation code described by
Wada \& Norman (2001) to obtain
density and temperature spatial distributions. The parameters for
the simulation are set as follows.
The velocity dispersion derived from the line width is
roughly 100 km s$^{-1}$ (ranging from 40 to 300 km s$^{-1}$;
e.g., Prochaska \& Wolfe 1998). Therefore, we fix the circular
velocity as $v_{\rm c}=100$ km s$^{-1}$. We assume a
formation redshift $z_{\rm vir}=3$. By using the spherical
collapse model for galaxy formation (see equations 2 and 7 of
Hirashita \& Ferrara 2002), those values yield  a virial mass
of $M_{\rm vir}=8.0\times 10^{10}~M_\odot$
and a radius of the dark matter halo of $r_{\rm vir}=34$ kpc.
If the baryon fraction is assumed to be equal to
$\Omega_{\rm b}/\Omega_{\rm M}$, the gas mass contained in
the galactic disk within the halo is estimated to be
$M_{\rm disk}=1.1\times 10^{10}f_{\rm disk}~M_\odot$, where
$f_{\rm disk}$ is the gas fraction contained in the disk.
According to Navarro \& Steinmetz (2000), we estimate
$f_{\rm disk}\simeq 0.2$ for our cosmological parameters.
We do not include star formation and stellar feedback.
The UV radiation for the photoelectric heating is assumed to be 
1/100 of the local Galactic value. 
The following conclusions, however, are not affected by the
details of the local radiation field, because the density and temperature
distributions of the dense gas where \H2 molecules are mainly formed are
not very sensitive to the UV intensity. We have confirmed that the
dense regions survive even in the strongest UV intensity treated in
this Letter.

To solve our problem it is necessary to follow the evolution of
the system from galactic scales down to the small regions
where \H2 forms. We use a resolution of $2048\times 2048$
cells with a fixed cell size of 0.49 pc, so as to reach a
global 1 kpc scale simulation. The typical radius of the
gaseous disk can be estimated to be
$0.18r_{\rm vir}=6.1$ kpc
(e.g., Ferrara, Pettini \& Shchekinov 2000). With this size, the
column density of hydrogen nuclei in the vertical direction of the
disk is $\sim 1\times 10^{22}f_{\rm disk}$ cm$^{-2}$. We simulated
the central 450 pc radius of the disk, in which  we assume that a
fraction $(0.45/6.1)^2$ of $M_{\rm disk}$ is contained. Then, we
simulate an exponential disk with a scale length of 100 pc, to 
check the effects of radial density profiles. From a series
of tests, we have been able to assess that 
the following results are not affected
significantly by the specific profile assumption.

The whole timescale of the simulation is 60 Myr, when a
quasi-stationary density distribution function is achieved. This
timescale is larger than the rotation time in the simulated
region: $2\pi (450~{\rm pc})/100~{\rm km~s}^{-1}=30$ Myr. For the
cooling function, we assume a metallicity of 0.1 $Z_\odot$,
appropriate for DLAs.

\begin{figure*}
\begin{center}
\includegraphics[width=7.cm]{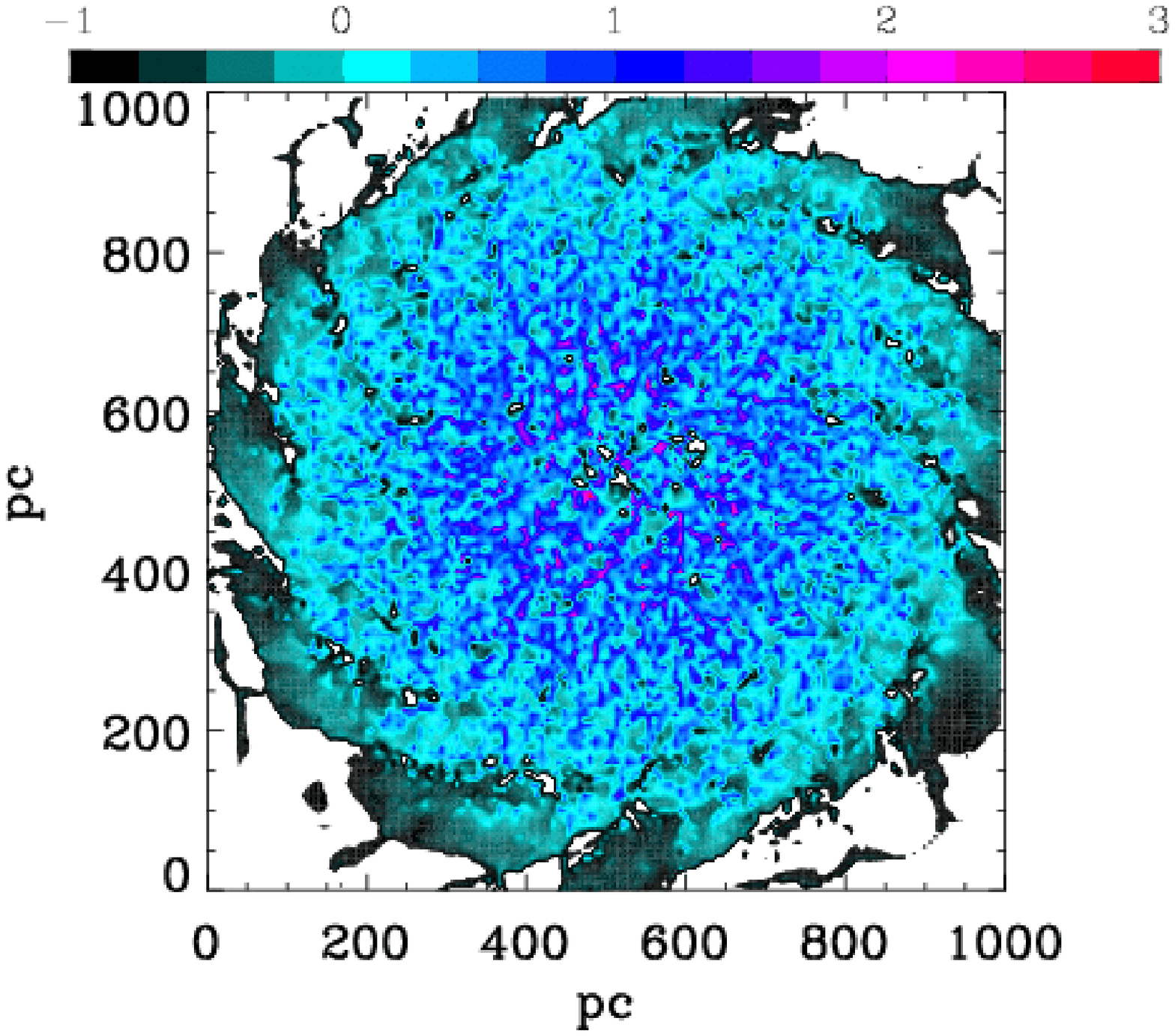}
\includegraphics[width=7.cm]{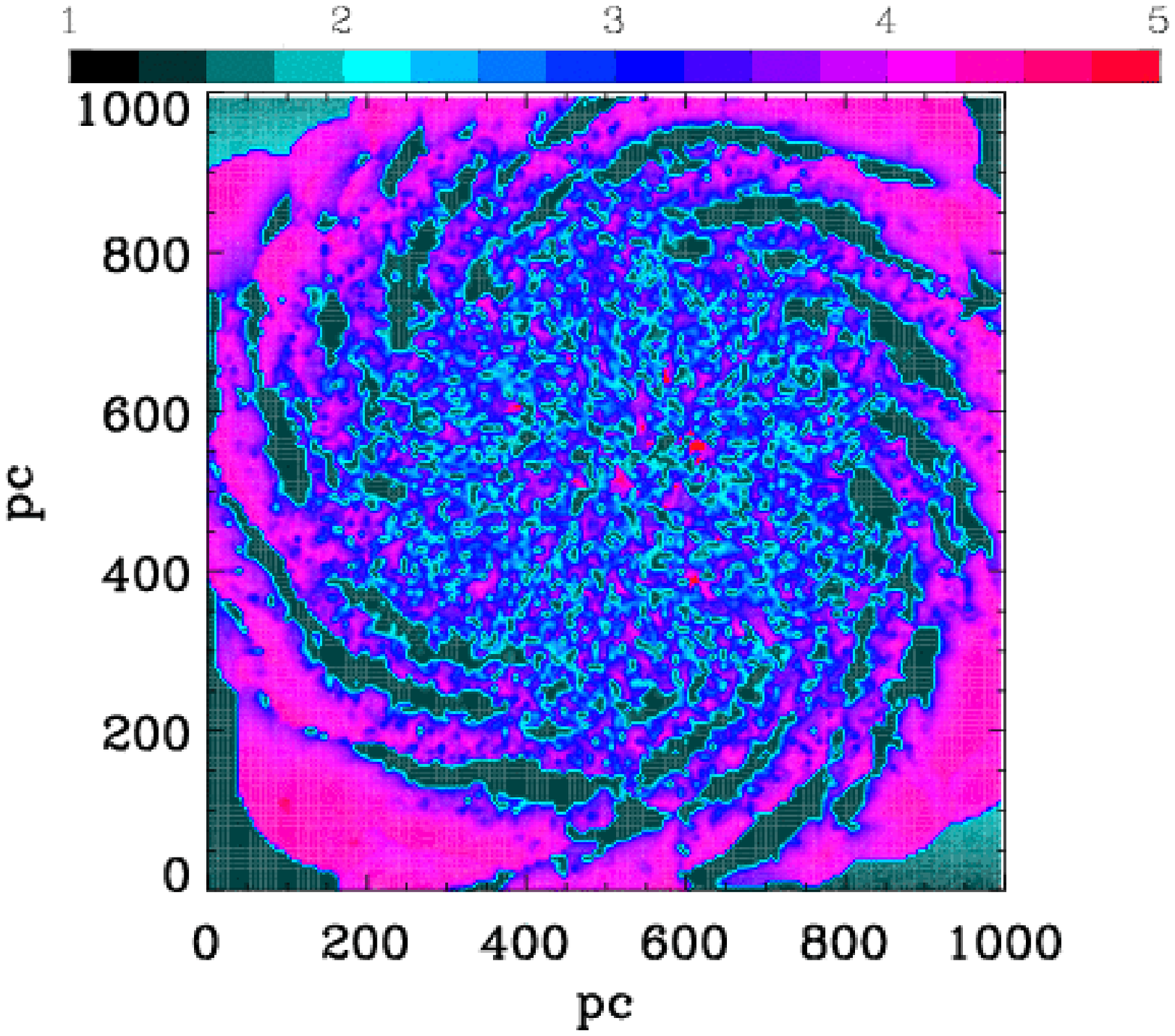}
\end{center}
\caption{
Spatial distribution of density (left) and
temperature (right) obtained from the simulation. The color bars
show the levels of $\log\nH$ and $\log T$, respectively.
\label{fig:temp_dens}}
\end{figure*}

In Fig.\ \ref{fig:temp_dens}, we show the density and temperature
distributions calculated by the simulation. A major part (86\%) of
the disk is covered with regions with $\nH <10~{\rm cm}^{-3}$
($\nH$ is the number density of hydrogen nuclei). This density
is roughly consistent with that derived observationally by
Silva \& Viegas (2002). The gas temperature $T$ is higher than
500 K in a large part of the disk (62\%), and indeed
Chengalur \& Kanekar (2000) observationally derived
$T>500$ K for a large part of their sample. There are also
a few DLAs detected with $T\sim 100$ K (e.g.\
Chengalur \& Kanekar 2000). For more detailed
statistical study, a complete set of
simulations that cover a range of mass, rotational velocity,
etc. is necessary. The above qualitative agreement, however, assures a good
basis on which we discuss the properties of DLAs. Those
density and temperature maps are used to estimate the spatial
distribution of molecular fraction in the following.

\section{\H2 spatial distribution}\label{sec:spatial}

\subsection{Estimate of molecular fraction}\label{subsec:chemi}

We estimate the molecular fraction ($f_{\rm H_2}$) at each
grid point. Since we are interested in the typical metallicity
of DLAs ($\sim 1/10~Z_\odot$; Vladilo 2002), the gas phase
formation of \H2 through the formation of H$^-$ is negligible
compared with the formation on dust (Liszt 2002). On the other
hand, the dominant mechanism for the destruction of \H2 is the
photodissociation under a UVB intensity.

For simplicity, we assume equilibrium between \H2 formation
and destruction. In order to examine if this assumption is
justified, we first estimate the formation timescale of \H2 as
$\fH2\nH /R_1\sim 10^4$ yr, where we adopt the reaction rate
$R_1\sim 4.6\times 10^{-16}$ cm$^{-3}$ s$^{-1}$ (defined in
equation \ref{eq:formation}), $\fH2\sim 10^{-6}$ (about the
experimental detection limit), and $\nH\sim 100$ cm$^{-3}$
(typical density for molecule-forming regions). In this Letter,
we are interested in structures down to 0.5 pc (grid size), and
the hydrodynamical timescale in such structures can be estimated
as 0.5 pc/10 km s$^{-1}\simeq 5\times 10^4$ yr. We are
particularly interested in the cold clouds where \H2 can form,
and in such a region, the typical hydrodynamical timescale is
$>10^5$ yr because of their
low sound speed. Thus, the chemical equilibrium can be
reasonably assumed for $\fH2\la 10^{-6}$
and $\nH\ga 100$ cm$^{-3}$. Since such an equilibrium
becomes a bad approximation for high $\fH2$, some of the
molecular rich clouds
with $\fH2\ga 10^{-4}$ will disappear in the course of
hydrodynamical evolution. In a diffuse gas with
$\nH\la 10$ cm$^{-3}$, the chemical equilibrium may not be
realised because
of a slow reaction rate. This suggests that \H2 does not form
so efficiently as estimated from the equilibrium assumption in a
diffuse medium. Moreover, such a diffuse medium has a temperature
$\ga 10^3$ K. With such a high temperature, hydrogen atoms
may not stick to dust efficiently. Therefore, $\fH2$ in diffuse
region may be overestimated in this Letter. Fortunately this
strengthen our conclusion on the lack of \H2 in a diffuse
medium. A more consistent treatment of \H2 reaction and
hydrodynamics will be tackled in the future.

We adopt the formation rate of \H2 per unit volume and time,
$R_1$, by Hollenbach \& McKee (1979) (see also
Hirashita, Hunt, \& Ferrara 2002):
\begin{eqnarray}
R_1 & \hspace{-4mm}= & \hspace{-4mm}0.5\nH (1-f_{\rm H_2})n_{\rm d}
\pi a^2\bar{v}S_{\rm d}(T)\nonumber\\
& \hspace{-4mm}\simeq & \hspace{-4mm}4.6\times 10^{-16}S_{\rm d}(T)
\left(\frac{a}{0.1~\mu{\rm m}}\right)^{-1}\left(
\frac{{\cal D}}{10^{-3}}
\right)\nonumber\\
& \times & \hspace{-3mm}\left(
\frac{\nH}{10^2~{\rm cm^{-3}}}\right)^2\left(\frac{T}{10^2~{\rm K}}
\right)^{1/2}\left(\frac{\delta}{3~{\rm g}~{\rm cm}^{-3}}\right)
(1-\fH2 )\nonumber\\
& & \hspace{4.3cm}{\rm cm^{-3}~s^{-1}}\, ,
\label{eq:formation}
\end{eqnarray}
where $n_{\rm d}$ is the number density of grains, $a$ is the
radius of a grain (assumed to be spherical with a radius of
0.1 $\mu$m),
$\bar{v}$ is the mean thermal speed of hydrogen, ${\cal D}$ is
the dust-to-gas mass ratio, $\delta$
is the grain material density (assumed to be 3 g cm$^{-3}$), and
$S_{\rm d}(T)$ is the sticking coefficient of hydrogen atoms onto
dust. In equation (\ref{eq:formation}) we have substituted the dust
number density $n_{\rm d}$ with the dust-to-gas ratio ${\cal D}$ by
using
\begin{eqnarray}
n_{\rm d}\frac{4\pi}{3}a^3\delta =\nH m_{\rm H}{\cal D}\, ,
\end{eqnarray}
where $m_{\rm H}$ is the mass of a hydrogen atom. The sticking
coefficient is given by (Omukai 2000)
\begin{eqnarray}
S_{\rm d}(T) & \hspace{-4mm}= & \hspace{-4mm}[1+0.4(T+T_{\rm d})^{0.5}
+2\times 10^{-3}T+8\times 10^{-6}T^2]^{-1}\nonumber \\
& & \times [1+\exp(7.5\times 10^2(1/75-1/T_{\rm d}))
]^{-1}\, ,
\end{eqnarray}
where $T_{\rm d}$ is the dust temperature. In this Letter,
$T_{\rm d}$ is assumed to be 20 K (a typical temperature under
the local interstellar radiation field), but the following
result is insensitive to this value as long as $T_{\rm d}\la 70$ K.

The photodissociation rate in units of cm$^{-3}$ s$^{-1}$, $R_2$,
is estimated by (Abel et al.\ 1997)
\begin{eqnarray}
R_2=(4\pi )\, 1.1\times 10^{-13}n_{\rm H_2}J_{21}\,
S_{\rm shield}[N({\rm H_2}),\, N({\rm dust})]\, ,
\label{eq:dissociation}
\end{eqnarray}
where $n_{\rm H_2}$ (cm$^{-3}$) is the number density of \H2,
$J_{21}$ ($10^{-21}$ erg s$^{-1}$ cm$^{-2}$ Hz$^{-1}$ str$^{-1}$)
is the UVB intensity at the Lyman limit wavelength (912 \AA)
averaged for all the solid angle,
$S_{\rm shield}(N({\rm H_2}),\, N({\rm dust}))$ is the
correction factor of the reaction rate for \H2 self-shielding
and dust extinction. We adopt the correction for the \H2
self-shielding by Draine \& Bertoldi (1996) (see also
Hirashita \& Ferrara 2002). Then, we estimate
$S_{\rm shield}$ as
\begin{eqnarray}
S_{\rm shield}={\rm min}\left[1,\, \left(
\frac{\nH\fH2 H}{10^{14}~{\rm cm}^{-2}}\right)^{-0.75}\right] e^{
-\pi a^2n_{\rm d}H}\, ,
\end{eqnarray}
where $H$ is the typical thickness of the disk which is assumed to
be 100 pc in this Letter.

In fact, in some DLAs with \H2 detection, the \H2 excitation is
consistent with the radiation field comparable to the local
Galactic radiation field (Ge \& Bechtold 1997;
Ledoux, Srianand, \& Petitjean 2002;
Petitjean, Srianand, \& Ledoux 2002).
In this case, the photo-dissociation rate can be much higher.
However, the \H2-detected DLAs could be biased to star-forming
molecular regions, and it is still unknown if DLAs in general 
are exposed to such a strong radiation field. Thus, we concentrate on the
UVB, which is common for all the DLAs. The equilibrium condition
$R_1=R_2$ with temperature and density at each grid point gives
$\fH2$ for each point. Thus, we obtain the spatial distribution of
$\fH2$ under a certain set of $({\cal D},\, J_{21})$.

\subsection{Molecular fraction maps}\label{subsec:map}

The metallicity level of DLAs ($\sim 0.1~Z_\odot$) implies that
the dust-to-gas ratio of DLAs is typically 10\% of the
Galactic (Milky Way) value. In this Letter, we assume the
Galactic dust-to-gas ratio to be 0.01. Therefore, we examine the
dust-to-gas ratio around 0.001. The UVB intensity is considered
to be $\j21\sim 0.3$--1 around $z\sim 3$
(Giallongo et al.\ 1996;
Cooke, Espey, \& Carswell 1997; Scott et al.\ 2000;
Bianchi, Cristiani, \& Kim 2001). Since the UVB intensity is
likely to be lower at $z\la 1$ (e.g.\ Scott et al.\ 2002), we
examine $\j21\sim 0.01$--1 in this Letter.

\begin{figure}
\begin{center}
\includegraphics[width=6.5cm]{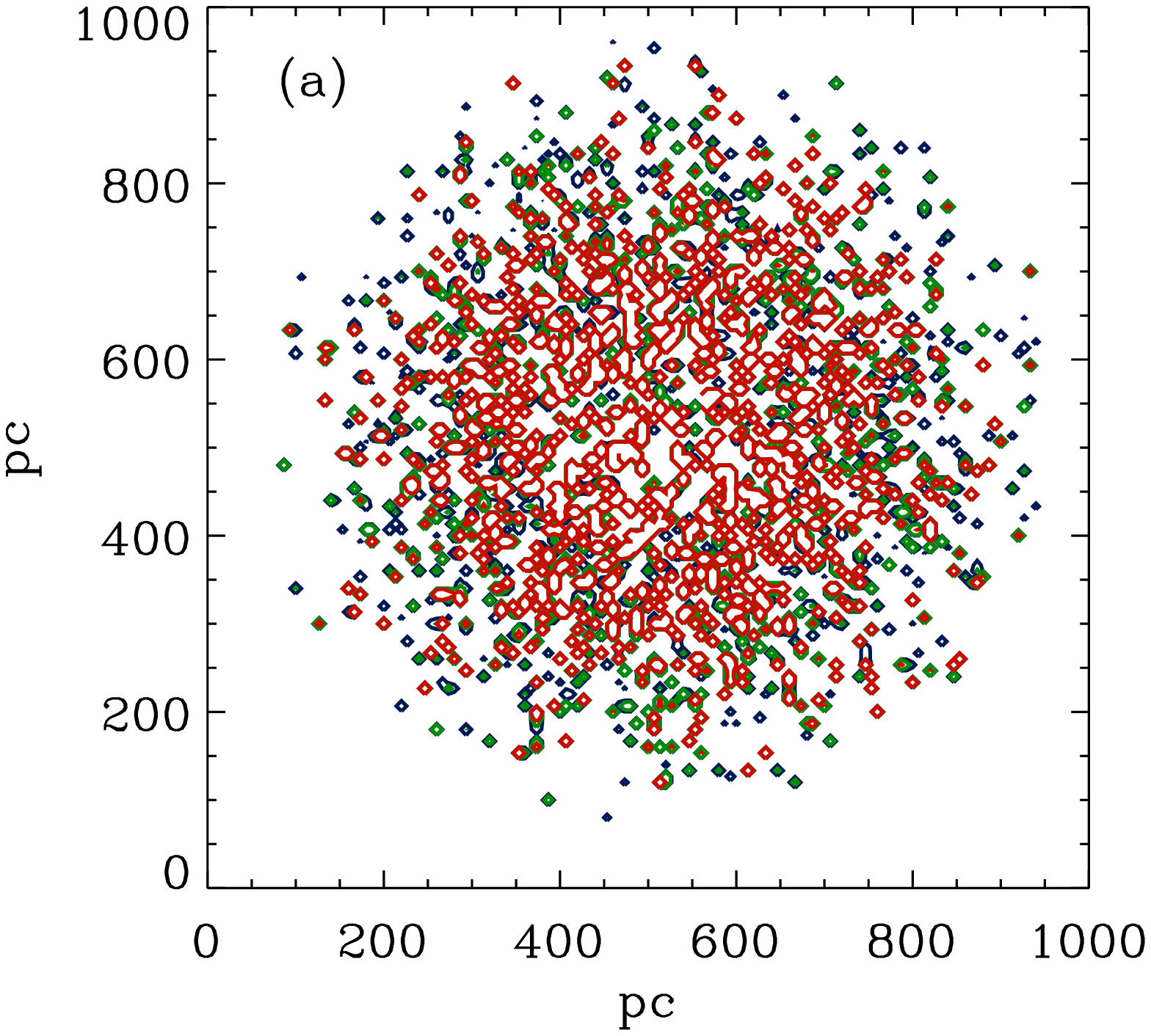}
\includegraphics[width=6.5cm]{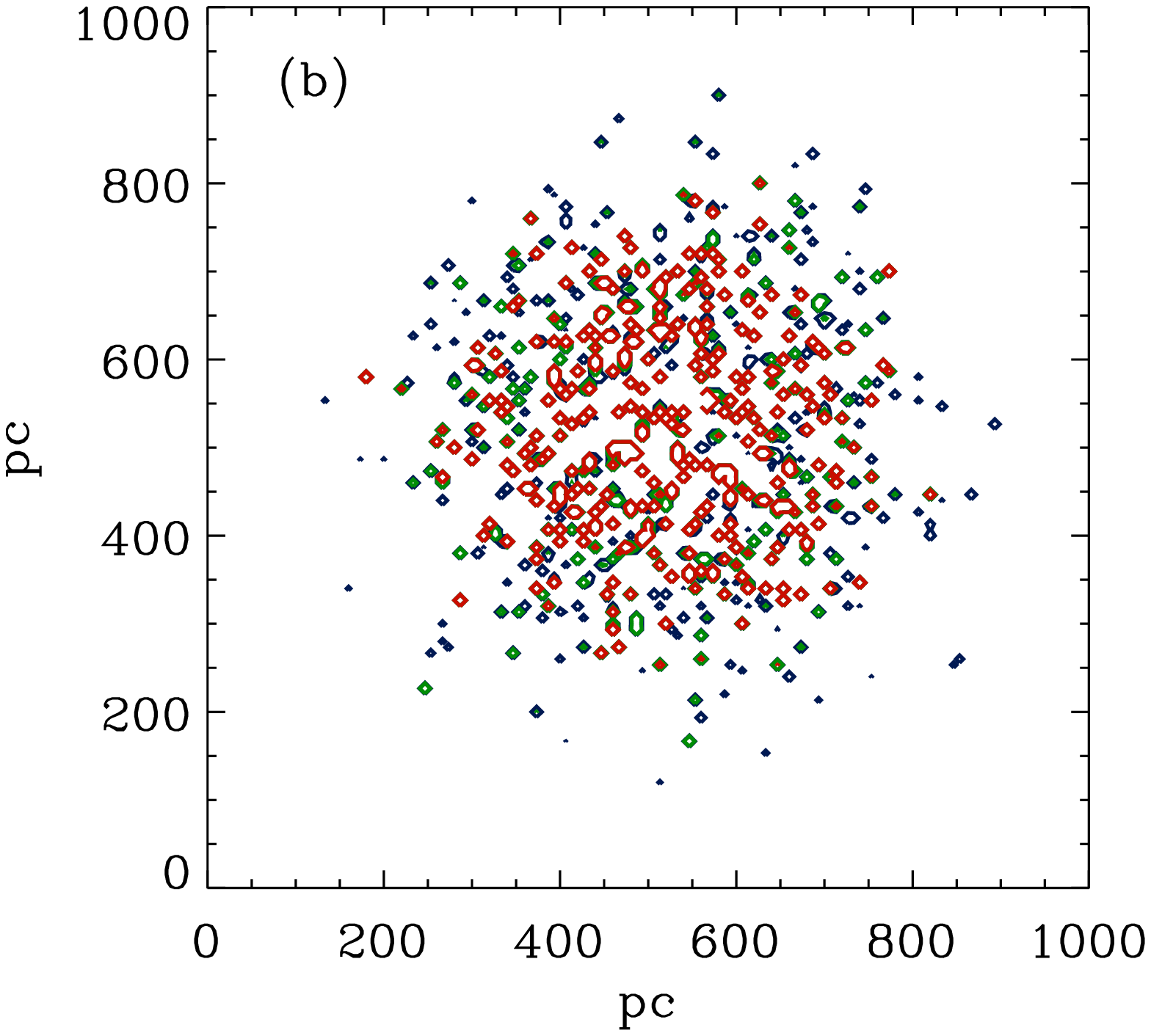}
\includegraphics[width=6.5cm]{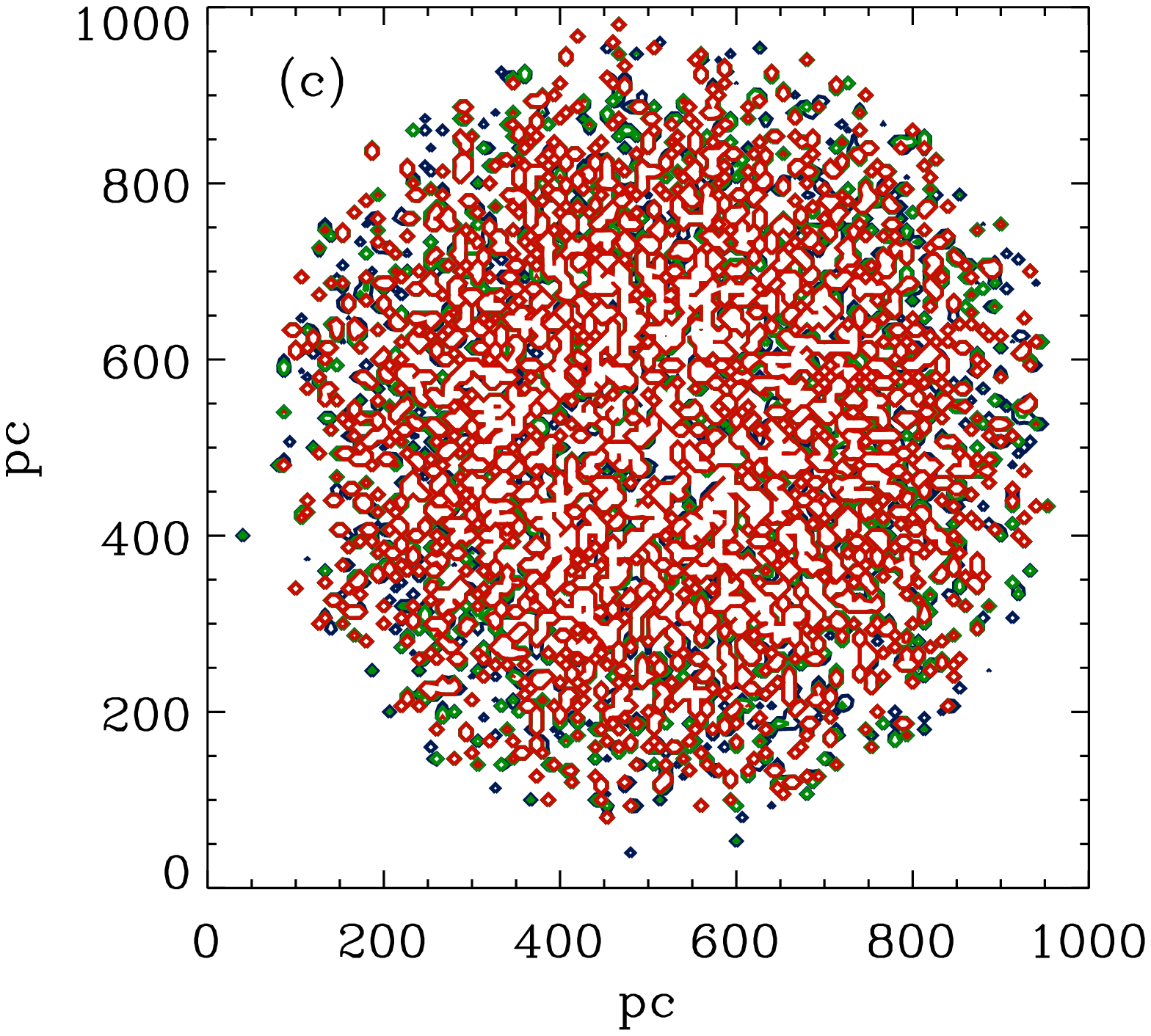}
\end{center}
\caption{Spatial distribution of molecular fraction ($\fH2$) for
(a) ${\cal D}=0.001$ (10\% of the Galactic dust-to-gas ratio)
and $J_{21}=0.1$, (b) ${\cal D}=0.001$ and $J_{21}=0.3$,
and (c) ${\cal D}=0.003$ and $J_{21}=0.1$. The contour levels
are for $\fH2 =10^{-6}$, $10^{-5}$, and $10^{-4}$ (blue,
green, and red, respectively).
\label{fig:mol_frac}}
\end{figure}

\begin{figure}
\includegraphics[width=6cm]{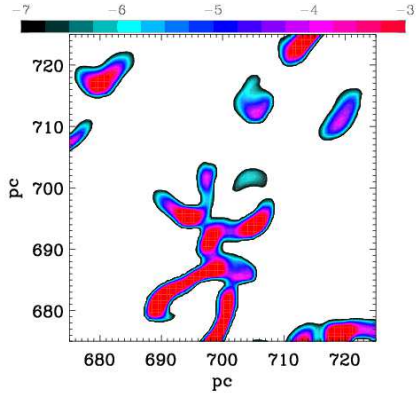}
\caption{Zooming on the region around (700 pc, 700 pc) in
Fig.\ \ref{fig:mol_frac}a. The color bars show
the levels of $\log\fH2$.\label{fig:mol_frac_zoom}}
\end{figure}

In Fig.\ \ref{fig:mol_frac}, we show the molecular fraction
($\fH2$) for (a) ${\cal D}=0.001$ (10\% of the Galactic
dust-to-gas ratio and $\j21 =0.3$, (b) ${\cal D}=0.001$ and
$\j21 =1$, and (c) ${\cal D}=0.03$ and $\j21 =0.1$. The contour
levels are for $\fH2 =10^{-6}$, $10^{-5}$, and $10^{-4}$.
In order to see the fine
structure of the molecular distribution, we also zoom on the
region around (700 pc, 700 pc) in Fig.\ \ref{fig:mol_frac_zoom}.
We find that molecular-rich regions are distributed very
inhomogeneously and concentrate in small ($\sim 10$ pc) clumps.

We define the covering fraction $C(>f)$ as the fraction of the
area where the
molecular fraction is larger than $f$ in the central 450 pc
($\sim$ radius of the simulated area) region. The disk is seen
face-on and projection effects of disk inclination is not
considered in this paper. Since
the disk thickness is much smaller than the size of the
galaxy, the effect of the inclination on the surface covering
fraction is expected to be small. Such a projection effect
can be quantitatively investigated in a future three-dimensional
simulation.

Since the typical detection limit for $\fH2$ is $10^{-6}$, we
are particularly interested in $C(>10^{-6})$. Observationally,
$C(>10^{-6})$ can be interpreted as the
probability of detecting a gas with $\fH2 >10^{-6}$ under a
condition that such a disk as simulated in this Letter hits the
line of sight to a distant quasar. If we assume that the
simulated disk is typical for DLAs, $C(>10^{-6})$
approximates the probability
of detecting H$_2$ in a given DLA with a level of
$\fH2\ga 10^{-6}$. In Table \ref{tab:mol_frac},
we show $C(>10^{-6})$ as a function of ${\cal D}$ and $\j21$. The
covering fraction is very
sensitive to both parameters in the range of interest. For
another observational thresholds, a power-law scaling
$C(>f)\propto f^{-\alpha}$ is applicable in
$10^{-7}\la f\la 10^{-4}$ with $\alpha =0.11$, 0.13, and 0.10, for
the same parameter sets as Figs.\ \ref{fig:mol_frac}a, b, and c,
respectively.

\begin{table}
\begin{center}
\caption{Covering fraction of the area with $\fH2 >10^{-6}$
(above the typical observational detection limit) in \% for
various ${\cal D}$ and $J_{21}$}
\begin{tabular}{ll|lllll}\hline
& ${\cal D}$ & 0.0001 & 0.0003 & 0.001 & 0.003 & 0.01 \\
$J_{21}$ & & & & & & \\ \hline
0.01 & & 3.8 &  15 &  34 & 57 &  85 \\
0.03 & & 1.7 &  8.5 & 25 & 46 &  72 \\
0.1  & & 0.75 & 3.7 & 16 &  34 & 60 \\
0.3  & & 0.43 & 1.8 & 9.5 &  25 & 48 \\
1    & & 0.26 & 0.81 & 4.9 & 16 & 37 \\
\hline
\end{tabular}
\label{tab:mol_frac}
\end{center}
\end{table}

\section{SUMMARY AND DISCUSSION}\label{sec:discussion}

Our results indicate that the fraction of the area with detectable
molecular fraction (typically $\fH2>10^{-6}$) is very small
for small dust-to-gas ratio (${\cal D}\la 0.001$, \ie
$\la 10$\% of the Galactic dust-to-gas ratio) and in a strong UVB
radiation ($\j21\ga 0.1$). Thus, it is rare that a molecular-rich
region hits the line of sight to a distant quasar. This means that
DLAs with a detectable molecular fraction are rare objects.

The lack of H$_2$ detection from DLAs does not necessarily mean
the lack of gas in molecular form. Under a strong UVB, molecules
are predominantly
confined in small areas. However, in such areas, the
molecular fraction is as high as $\ga 10^{-3}$.
This molecular fraction is large enough to provide the minimum
radiative cooling necessary to ignite the star formation 
process.

Table \ref{tab:mol_frac} shows that the covering fraction of
molecular-rich regions sensitively changes
as the dust-to-gas ratio and/or the UVB intensity vary. If the
dust-to-gas ratio is 10\% of the Galactic value and the UVB is as
strong as expected
at high redshift ($\j21\ga 0.1$), $C(>10^{-6})\la 10$\%; on the contrary
$C(>10^{-6})\ga 20$\% if the dust-to-gas ratio is $\ga 30$\% of the
Galactic value. Such sensitive dependence on ${\cal D}$ may
explain the observational correlation between molecular fraction and
the dust abundance for DLAs (Ge et al.\ 2001), similar to the local
Universe. This underlines the importance  of dust for
the \H2 chemical network at high redshift.

We also found that a detectable amount of \H2 is localised in regions
with $\nH\ga 100~{\rm cm}^{-3}$ and $T\la 100~{\rm K}$. Those
ranges of density and temperature are consistent with those
observationally derived for \H2-detected DLAs
(e.g., Ledoux et al.\ 2002). Although most of the DLAs may contain \H2,
molecular clouds are difficult to
detect in absorption if they are embedded in a dust-poor and/or
UV-intense environment,
because a large fraction of the area is covered by a diffuse, molecule-poor
medium. Indeed, most of DLAs are observationally suggested to arise
selectively in diffuse neutral gas not associating with a detectable
amount of \H2 (e.g.\ Petitjean et al.\ 2000). Therefore, the observed lack
of molecules from absorption measurements does not necessarily
indicates the lack of molecular clouds, but probably 
reflects the small size of molecular-rich regions.

\section*{Acknowledgments}
We thank the anonymous referee for helpful comments and
F. Sigward for helping with IDL programming.  HH is supported by JSPS
Postdoctoral Fellowship for Research Abroad.
PR is supported by the {\it Deutsche Forschungsgemeinschaft}.
Hydrodynamical simulations were carried out on Fujitsu VPP5000 at 
ADAC, NAOJ.

\end{document}